\documentclass[doublecol]{epl2} % for 2 columns style with line numbers
\usepackage{cite}
\usepackage[utf8]{inputenc}
\usepackage{graphicx}
\usepackage{graphicx,epstopdf}
\usepackage{dcolumn}
\usepackage{amsmath}
\usepackage{lineno,hyperref}
\usepackage{amsfonts}
\usepackage{amsthm}
\usepackage{bbm}
\usepackage{amssymb}
\usepackage{mathrsfs}
\PassOptionsToPackage{caption=false}{subfig}
\usepackage{subfig}
\usepackage{color}
\usepackage[normalem]{ulem}

\newcommand{\Dcal}{\mathcal{D}}

\newcommand{\Ccal}{\mathcal{C}}
\newcommand{\Ecal}{\mathcal{E}}

\newcommand{\Ucal}{\mathcal{U}}
\newcommand{\Vcal}{\mathcal{V}}

\usepackage{blindtext}
\usepackage{lipsum}
\usepackage{bbm}

\newcommand{\1}{\mathbbm{1}}

\newcommand{\ket}[1]{| #1 \rangle}

\newcommand{\interpro}[2]{\langle #1 | #2 \rangle}
\newcommand{\bra}[1]{\langle #1 |}

\newcommand{\tr}[1]{ \text{Tr}\left\{ #1 \right\}}

\title{Charging power and stability of always-on transitionless driven quantum batteries}

\author{Luiz F. C. Moraes$^\ast$\inst{1}, Andreia Saguia$^\dagger$\inst{1}, Alan C. Santos$^\ddagger$\inst{2}, Marcelo S. Sarandy$^\&$\inst{1}}
\shortauthor{Moraes, Saguia, Santos, Sarandy}

\institute{                    
	\inst{1} Instituto de F\'{i}sica, Universidade Federal Fluminense, Av. Gal. Milton Tavares de Souza s/n, Gragoat\'{a}, 24210-346 Niter\'{o}i, Rio de Janeiro, Brazil \\
	\inst{2} Departamento de F\'{i}sica, Universidade Federal de S\~ao Carlos, P.O. Box 676, 13565-905, S\~ao Carlos, S\~ao Paulo, Brazil\\
	$^\dagger$lfcmoraes@id.uff.br, $^\dagger$amen@if.uff.br, $^\ddagger$ac\_santos@df.ufscar.br, $^\&$msarandy@id.uff.br
}


\abstract{The storage and transfer of energy through quantum batteries are key elements in quantum networks. 
	Here, we propose a {charger} design based on transitionless quantum driving (TQD), which allows for inherent control over 
	the battery charging time, with the speed of charging coming at the cost of the internal energy available to implement the dynamics. 
	Moreover, the TQD-based {charger} is also shown to be locally stable, which means that the {charger} can be disconnected from {the quantum battery} ({QB}) at any time after the energy transfer to the {QB}, with no fully energy backflow to the {charger}. This provides 
	a highly charged {QB} in an always-on asymptotic regime. We illustrate the robustness of the {QB} charge against time 
	fluctuations and the full control over the evolution time for a feasible TQD-based {charger}. }

\begin{document}

\maketitle

%%%%%%%%%%%%%%%%%%%%%%%%%%%%%%%%%%%%%%%%%%%%%%%%%
\section{Introduction}
%%%%%%%%%%%%%%%%%%%%%%%%%%%%%%%%%%%%%%%%%%%%%%%%%

Quantum batteries (QBs) are quantum devices able to both temporarily store and then transfer 
energy~\cite{Alicki:13,PRL2013Huber,PRL2017Binder,PRL_Andolina}. They can potentially 
benefit from entanglement and other quantum correlations to charge faster than conventional 
classical batteries. QBs are potentially relevant as fuel for other quantum devices and, more 
generally, for boosting the development of quantum networks. A number of distinct 
experimental architectures have been proposed to implement QBs, such as spin systems~\cite{Le:18}, 
quantum cavities~\cite{Binder:15,Fusco:16,Zhang:18,Ferraro:18}, superconducting transmon 
qubits~\cite{Santos:19-a}, quantum oscillators~\cite{Andolina:18,Andolina:19} and spin batteries in 
quantum dots~\cite{Long:03}. 

A key challenge to implement QBs is to ensure the stability of the {charging and} energy transfer process. A QB is 
defined as stable if it can {be charged, with respect to a reference state, with no energy backflow (spontaneous discharge) at future times}. A solution for the stability problem 
based on the adiabatic theorem has been introduced in Ref.~\cite{Santos:20c}. As a first 
step, it is introduced the energy current (EC) operator, whose expectation value quantifies the rate of internal 
energy transferred from the {charger} to the {QB}. By taking the {QB} in pure states both at initial and final 
evolution times, the expectation value of the EC operator at the final time will be the 
time derivative of the extractable work from the {QB}~\cite{Santos:20c}. 
Then, it has been shown that the stability is achieved at the adiabatic time scale $\tau_a$ by 
initially preparing the {charger}-{QB} system in an eigenstate $|p_0\rangle$ of the total Hamiltonian acting on the 
{charger}-{QB} Hilbert space. Under such a condition, the expectation value of the 
EC operator adiabatically vanishes, preventing energy backflow to the {charger}~\cite{Santos:20c}. Therefore,  
the QB can be manufactured {to charge or discharge its stored energy} at any fixed total time $\tau_c$ such that 
$\tau_c > \tau_a$. Even though this provides a general adiabatic-based approach to yield stable QBs, which 
is independent of architectures and corresponding Hamiltonians, it turns out to be still a partial 
solution for the stability problem. First of all, it requires the validity of the adiabatic approximation, which 
imposes constraints on the energy gap dynamics and makes the system potentially more susceptible to 
decoherence due to the competition between the adiabatic and the relaxation time scales. Moreover, 
Ref.~\cite{Santos:20c} solves a particular stability problem, which we define here as the {\it global} stability problem.  
This means that stability is ensured by manufacturing the QB to complete the {charge} at a specific global time $\tau_c$ (such that $\tau_c > \tau_a$). 
The choice of $\tau_c$ is arbitrary as long as it is greater than $\tau_a$. However, the Hamiltonian dynamics 
is engineered to occur from the initial time $t=0$ up to $t=\tau_c$, with no time fluctuation. Therefore, 
global stability does not account for a possible energy backflow if the {charger} is still connected to the {QB} after 
the time $\tau_c$. The ability of keeping the {QB} charged with no energy revivals in the {charger} in an {\it always-on} asymptotic 
regime is what we define as {\it local} stability, which is not dealt with in the solution proposed in 
Ref.~\cite{Santos:20c}.

In this work, we propose a solution to the adiabatic time constraint as well as the local stability problem. 
The adiabatic hypothesis can be relaxed here by implementing transtionless quantum driving 
(TQD)~\cite{Demirplak:03,Demirplak:05,Berry:09}. TQD is a useful technique to mimic adiabatic 
quantum tasks at arbitrary finite time, providing a shortcut to adiabaticity. 
It has been applied for speeding up quantum gate Hamiltonians~\cite{Santos:15,Santos:16,Coulamy:16,Hegade:20},
heat engines in quantum thermodynamics~\cite{Adolfo:16},
quantum information processing \cite{Marcela:14,Xia:16,Chen:10}, among
other applications (see e.g., Ref.~\cite{Odelin:19} and references there in).
Naturally, there is a commitment between the speed of the evolution and the internal energy required to drive the dynamics. 
Moreover, the implementation of TQD usually requires many-body interaction terms in the Hamiltonian, whose number may 
exponentially grow with the size of the system. As we will show, 
the Hamiltonian complexity is tractable here, requiring up to three-body interactions. 
Concerning the local stability problem, we solve it through a suitable non-linear Hamiltonian interpolation, which yields local stability 
either in the adiabatic or in the TQD scenario. These non-linear interpolations keep a non-vanishing gap throughout the 
evolution while ensuring an asymptotically fueled {QB}. The extractable energy will be considered 
from the point of view of the ergotropy of the {QB}, which turns out to show robustness against time variation.

%%%%%%%%%%%%%%%%%%%%%%%%%%%%%%%%%%%%%%%%%%%%%%%%%
\section{Ergotropy and QB stability}
%%%%%%%%%%%%%%%%%%%%%%%%%%%%%%%%%%%%%%%%%%%%%%%%%

Let us begin by discussing some preliminary concepts, namely, the work extractable from a {charger} and the distinct kinds of 
stability behaviors in the charging and discharging energy transfer processes. 

%%%%%%%%%%%%%%%%%%%%%%%%%%%%%%%%%%%%%%%%%%%%%%%%%
\subsection{Ergotropy and internal energy}
%%%%%%%%%%%%%%%%%%%%%%%%%%%%%%%%%%%%%%%%%%%%%%%%%

The usefulness of the energy tranferred by a { charger} can be quantified in terms of the \textit{ergotropy}, which 
is defined as the maximum energy that can be extractable as work from a quantum system through unitary 
operations $V$. Ergotropy has been proposed  by Allahverdyan {\it {et al.}}~\cite{Allahverdyan:04}, reading
\begin{align}
\Ecal = \tr{\rho H_{0}} - \min_{V\in \Vcal}\tr{V\rho V^\dagger H_{0}} , \label{Erg1}
\end{align}
where $\rho$ denotes the { charger} state, $V$ is a unitary operator belonging to the set $ \Vcal$ of all possible 
unitary operations over the system, and $H_{0}$ is the reference Hamiltonian that defines the internal 
battery structure (defining the { charger} energy states). Without loss of generality, we can write $H_{0}$ by using its 
spectral decomposition $H_{0}\! =\!\sum\nolimits_{n=1}^{\Dcal} \epsilon_{n} \ket{\epsilon_{n}}\bra{\epsilon_{n}}$, 
with $\Dcal$ being the dimension of the Hilbert space and $\epsilon_{1}\!\leq\!\epsilon_{2}\!\leq\!\cdots\!\leq\! \epsilon_{\Dcal}$. 
Therefore, the empty battery energy is taken as $\epsilon_{1}$ and the full energy value as $\epsilon_{\Dcal}$, 
so that the { charger} has an energy capacity given by $\Ecal_{\text{max}}\!=\!\epsilon_{\Dcal} - \epsilon_{1}$. 
Let us now consider the spectral decomposition of the { charger} density operator 
$\rho\! =\!\sum\nolimits_{n=1}^{\Dcal} \varrho_{n} \ket{\varrho_{n}}\bra{\varrho_{n}}$, where $\ket{\varrho_{n}}$ are 
eigenvectors of $\rho$ with eigenvalues $\varrho_{n}$, so that $\varrho_{1}\!\geq\!\varrho_{2}\!\geq\!\cdots\!\geq\! \varrho_{\Dcal}$. 
By taking this arrangement order, the minimization in Eq.~(\ref{Erg1}) is attained for 
$V=\sum\nolimits_n  \ket{\epsilon_{n}}\bra{\varrho_{n}}$, yielding  
\begin{equation}
\Ucal_0 = \min_{V\in \Vcal}\tr{V\rho V^\dagger H_{0}} = \sum_{n=1}^{\Dcal}  \varrho_{n} \epsilon_{n}. \label{Emin}
\end{equation}
Notice that $\Ucal_0$ arises from the operation $V$ that drives $\rho$ to a diagonal form in the energy eigenbasis. 
Then, Eq.~\eqref{Erg1} can be rewritten in a more convenient way as~\cite{Allahverdyan:04,Baris:20}
\begin{align}
\Ecal = \sum_{i=1}^{\Dcal} \sum_{n=1}^{\Dcal} \varrho_{n}\epsilon_{i} \left( |\interpro{\varrho_{n}}{\epsilon_{i}}|^2 - \delta_{ni} \right) . 
\label{Ergo2}
\end{align}
From Eq.~(\ref{Ergo2}), it is worth emphasizing that not all the initial internal energy can be effectively transferred as work 
(see, e.g., Refs.~\cite{Kamin:20-1,James:20}). In fact, by writing the internal energy $\Ucal (\rho,H_0)$ for the reference 
Hamiltonian $H_{0}$, we have 
\begin{align}
\Ucal (\rho,H_0) = \tr{\rho H_{0}} =  \sum_{i=1}^{\Dcal} \sum_{n=1}^{\Dcal} \varrho_{i}\epsilon_{n}|\interpro{\varrho_{n}}{\epsilon_{i}}|^2 ,
\end{align}
which means that 
\begin{align}
\Ecal = \Ucal({\rho,H_{0}}) - \Ucal_{0}. \label{Ergotropy}
\end{align}
Therefore, since ergotropy is positive for any process (by construction), the internal energy of the system 
needs to satisfy $\Ucal({\rho,H_{0}})\!>\!\Ucal_{0}$ for an amount of extractable work to be available in the QB setup. 

%%%%%%%%%%%%%%%%%%%%%%%%%%%%%%%%%%%%%%%%%%%%%%%%%
\subsection{Stability criteria}
%%%%%%%%%%%%%%%%%%%%%%%%%%%%%%%%%%%%%%%%%%%%%%%%%

A quantum system exhibiting discrete constant eigenvalues obeys the quantum version of the recurrence theorem of 
Poincar\'e~\cite{Bocchieri:57}, which establishes that, if $\ket{\psi(t_0)}$ is the system state vector at time $t_0$ and 
$\epsilon$ is any positive number, then a future time $T$ will exist such that 
$\|\, \ket{\psi(T)} - \ket{\psi(t_0)}\,\|^2\!<\!\epsilon$, with $\| \, \ket{\phi} \, \|$ denoting the vector norm 
$\| \, \ket{\phi} \, \|\!=\!\sqrt{\langle \phi | \phi \rangle}$. Therefore, any QB obeying the quantum recurrence theorem will 
be unstable, {\it{\i.e.}} subjected to spontaneous charging and discharging processes~\cite{Santos:19-a,Santos:20c,James:20}. 
As a strategy to avoid spontaneous energy transfer between {a charger} and the {QB}, we may consider suitable 
time-dependent evolutions, such as the adiabatic dynamics~\cite{Born:28,Kato:50}. Indeed, this leads to the 
possibility of designing stable QBs~\cite{Santos:19-a,Santos:20c}. 

In this scenario, Ref.~\cite{Santos:20c} introduced an adiabatic-based solution for a specific stability problem, 
which we define here as the {\it global} stability problem.  Global stability requires the QB to complete the {charge}
at a freely programmed global time $\tau_c$, which is required to be large enough to obey the adiabatic approximation~\cite{Messiah:book,Sarandy:04}. 
The dynamics is then assumed to {\it globally} occur from the initial time $t=0$ up to $t=\tau_c$, with absolutely 
no time fluctuation. Therefore, global stability does not account for a possible energy backflow if the {charger} is still 
connected to the {QB} after the time $\tau_c$. 

In order to prevent energy revivals in the {charger} in the {\it always-on} time regime, we also introduce here the concept of {\it local} 
stability. By taking $\epsilon$ as a positive number, local stability is defined through the requirement that the 
{QB} ergotropy $\Ecal_{{QB}}(t)$ is such that $\Ecal_{{QB}}(t) > \epsilon$ after a finite charging time $\tau_c$. 
Notice then that 
local stability refers to the prevention of full energy backflow to the {charger} {\it locally} in time, 
as it would be expected from the quantum recurrence theorem. 
Naturally, the large the value of $\epsilon$ the higher the performance of the locally stable QB. 
Let us consider that systematic errors or an inherent longer turned-on {charger}-{QB} connection 
could lead us to an evolution time $\Delta t_{\text{real}}\!=\!\tau_c + \Delta t$. Then, 
the battery ergotropy can deviate from its maximum value as 
$\Ecal_{\text{real}}(\Delta t)\!=\! \Ecal_{\text{max}} - |\Delta \Ecal_{\text{dev}}(\Delta t)|$. 
By rearranging this expression, we have
\begin{equation}
\Ecal_{\text{real}}(\Delta t)/\Ecal_{\text{max}} = 1 - \eta_{\text{ls}}(\Delta t) ,
\label{etalc}
\end{equation}
where $\eta_{\text{ls}}(\Delta t)\equiv |\Delta \Ecal_{\text{dev}}(\Delta t)| / \Ecal_{\text{max}}\!\in\![0,1]$ can be 
understood as a \textit{local stability coefficient} for the QB. There are situations in 
which the charge $\Ecal_{\text{max}}$ is not achieved, but we have an asymptotic amount of ergotropy 
$\Ecal_{\text{asy}}\!<\!\Ecal_{\text{max}}$~\cite{James:20}. In this case, Eq.~(\ref{etalc}) can be generalized 
by replacing $\Ecal_{\text{max}}$ for $\Ecal_{\text{asy}}$ ($\Ecal_{\text{max}}$ is merely a chosen normalization). 
As an example, consider the case where one can use dark 
states to stabilize open system QBs~\cite{James:20}. Due to the effect of decoherence, the maximum 
ergotropy is not achieved. However, we have an asymptotic value $\Ecal_{\text{asy}}\!>\!0$ for which,  
after some instant $t\!=\!\tau_{\text{c}}$, one gets $\eta_{\text{ls}}(\Delta t)\!=\!0$, for any $\Delta t\!>\!0$, 
since $\Delta \Ecal_{\text{dev}}(\Delta t>0)\!=\!0$. In this situation, one has an optimal locally  
stable charging process. It is important mentioning that there are situations for which stability is 
achieved only for some finite interval $\Delta t$, such as recently reported in non-Markovian~\cite{Kamin:20-1} 
and Floquet engineered QBs~\cite{Bai:20}.

%%%%%%%%%%%%%%%%%%%%%%%%%%%%%%%%%%%%%%%%%%%%%%%%%
\section{Stability of the adiabatic ergotropy transfer}
%%%%%%%%%%%%%%%%%%%%%%%%%%%%%%%%%%%%%%%%%%%%%%%%%

In this section, we will discuss the energy transfer process of QBs driven by adiabatic Hamiltonians. 
%For simplicity, we consider the total Hamiltonian in the interaction picture. Then, the internal battery 
% contribution $H_{0}$ does not develop any role on the battery charging power~\cite{Santos:20c}.
Let us consider the spectral decomposition of a time-dependent Hamiltonian 
{$H_{\text{ad}}(t) = \sum\nolimits_{n=1}^{\Dcal} E_{n} (t)\ket{n(t)}\bra{n(t)}$}.  
We prepare the initial state $\ket{\psi(0)}$ of the system in a single non-degenerate eigenstate 
{$\ket{n(0)}$} of $H_{\text{ad}}(0)$. By assuming an adiabatic evolution, the evolved state reads 
{$\ket{\psi(t)}\!=\! e^{i\theta_{\text{ad}}(t)}\ket{n(t)}$}, where the adiabatic phase is 
{$\theta_{\text{ad}}(t) \!=\! -E_{n}(t)/\hbar + i \interpro{n(t)}{\dot{n}(t)}$}. 
The density matrix is then {$\rho_{\text{ad}}(t)\!=\!\ket{n(t)}\bra{n(t)}$}, with no contribution 
of the adiabatic phase. %Thus, from Eq.~\eqref{Ergo2}, we obtain the ergotropy{
%\begin{equation}
%\Ecal_{\text{ad}}(t) = \sum_{i=1}^{\Dcal} \left( \epsilon_{i}(t) \, |\interpro{\epsilon_n(t)}{\epsilon_{i}(t)}|^2 \right) - \epsilon_{n}(t) .\label{ergoad}
%\end{equation}
%where $\ket{e_n}$ is the basis of the reference Hamiltonian of the QB, given by $H_{0}\!=\!\hbar\omega \sigma_{z}^{(\text{QB})}$} Eq.~(\ref{ergoad}) is valid for pure states. Otherwise, the ergotropy needs to be computed from its general formulation as in Eq.~\eqref{Erg1}.

The {charger} model adopted here is composed of a two-qubit cell (allowing for entanglement), whose internal structure is defined by the Hamiltonian 
$H_{0}\!=\!\hbar\omega ( \sigma_{z}^{(1)} + \sigma_{z}^{(2)} )$, with energy basis given by $\sigma_{z}\ket{n}\!=\!(-1)^{1-n}\ket{n}$  for $n \in \{0,1\}$. 
Under this configuration, each qubit has a storable energy capacity $\Ecal_{\text{max}}^{\text{qubit}}\!=\!2\hbar\omega$, with the corresponding 
full charge state $\ket{1}$.  
Consequently, one can define the {charger} empty and full charge states as $\ket{\text{emp}}\!=\!\ket{00}$ and $\ket{\text{full}}\!=\!\ket{11}$, respectively, 
so that for the {charger} in the state $\ket{\text{full}}$ the ergotropy reads $\Ecal_{\text{max}}^{\text{cell}}\!=\!2\Ecal_{\text{max}}^{\text{qubit}}\!=\!4\hbar\omega$. 
The initial {charger} state is taken as in Ref.~\cite{Santos:20c}, which is given by the singlet state
\begin{align}
\ket{\psi(0)}_{{\Ccal}} = \frac{\ket{0}_{{\Ccal}_1}\ket{1}_{{\Ccal}_2} - \ket{1}_{{\Ccal}_1}\ket{0}_{{\Ccal}_2}}{\sqrt{2}} ,
\label{singlet}
\end{align}
with ${\Ccal}_i$ denoting the $i$-th qubit of the {charger} ($i \in \{1,2\}$). The maximally entangled state in Eq.~(\ref{singlet}) is responsible 
for the quantumness of the {charger}, allowing for a more efficient energy transfer to a {QB} than a factorizable state and for the design of {switchable 
charger}s~\cite{Santos:20c}. The process we will specifically consider here is the ergotropy transfer from the {charger} to  the {QB}. The {QB} will be taken here a single 
qubit system, initially in the state $\ket{0}_{\text{{QB}}}$. Then, after some time $\tau_{\text{c}}$, we expect to transfer the ergotropy initially stored in 
the {charger} to the {QB}. By denoting ${\Ccal}\!=\!{\Ccal}_{1}\otimes {\Ccal}_{2}$ as the {charger} Hilbert space, the dynamics is then expected to lead the system from 
the initial state $\ket{\phi(0)}\!=\!\ket{\psi(0)}_{{\Ccal}}\ket{0}_{\text{{QB}}}$ to the final state 
$\ket{\phi(\tau_{\text{c}})}\!=\!\ket{0}_{{\Ccal}_1}\ket{0}_{{\Ccal}_2}\ket{1}_{\text{{QB}}}$ through a stable charging process of the {QB}.
To this aim, we drive the system by a piece-wise time-dependent Hamiltonian given by
\begin{align}
H_{\text{ad}}(t) = \left[1 - f(t)\right] H_{\text{ini}} + f(t)\left[1 - f(t)\right] H_{\text{inter}} + g(t)H_{\text{fin}} , \label{AdHamil}
\end{align}
where $f(t)$ and $g(t)$ are real functions satisfying $f(0)\!=\!g(0)\!=\!0$ and $f(\tau)\!=\!g(\tau)\!=\!1$, for some total evolution time $\tau$, 
and each time-independent Hamiltonian reads
\begin{align}
H_{\text{ini}}&=\hbar\Omega \left[ \sigma_{x}^{({\Ccal}_1)}\sigma_{x}^{({\Ccal}_2)} + \sigma_{y}^{({\Ccal}_1)}\sigma_{y}^{({\Ccal}_2)} \right] \1_{\text{{QB}}} ,\\
H_{\text{inter}}&=\hbar\Omega \1^{({\Ccal}_1)}\left[ \sigma_{x}^{({\Ccal}_2)}\sigma_{x}^{\text{{QB}}} + \sigma_{y}^{({\Ccal}_2)}\sigma_{y}^{\text{{QB}}} \right] ,\\
H_{\text{fin}}&=\hbar\Omega \left[ \sigma_{z}^{({\Ccal}_1)}\1^{({\Ccal}_2)}\sigma_{z}^{\text{{QB}}} + \1^{({\Ccal}_1)}\sigma_{z}^{({\Ccal}_2)}\sigma_{z}^{\text{{QB}}} \right] .
\end{align}

\begin{figure}[t!]
	\centering
	\includegraphics[scale=0.32]{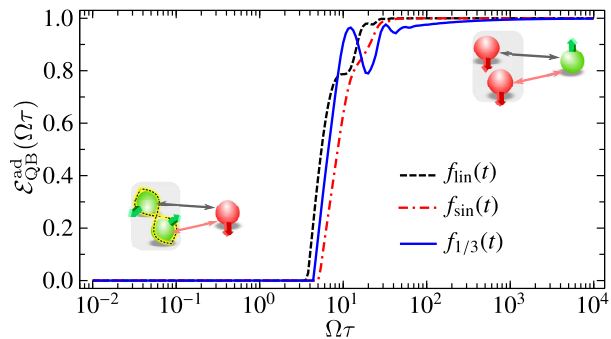}
	\caption{Ergotropy $\Ecal^{\text{ad}}_{\text{{QB}}}(\tau)$ {(as a multiple of $\Ecal_{\text{max}}^{\text{qubit}}$)} of the {QB} at the end of the adiabatic evolution (at time $t\!=\!\tau$). 
		Global stability is achieved for $\tau$ at the adiabatic scale. }
	\label{Fig-ErgoAdiab-Global}
\end{figure}

Each contribution to the Hamiltonian plays a specific role in the energy transfer process. $H_{\text{ini}}$ allows us to encode the state 
$\ket{\phi(0)}$ as one of its eigenstates. The final desired state $\ket{\phi(\tau_{\text{c}})}$ can then be correctly addressed because 
it is an eigenstate of $H_{\text{fin}}$ with identical energy and parity as the initial state~\cite{Santos:20c}. Concerning $H_{\text{inter}}$, it is 
introduced here in order to ensure a non-crossing ground state energy gap.

It is worth mentioning that our choice of the function $f(t)$ may affect the ergotropy transfer rate and the stability of the battery, since the minimum 
instantaneous gap depends on the function $f(t)$. As an illustration, we will consider three different functions given by 
$f_{\text{lin}}(t)\!=\!g_{\text{lin}}(t)\!=\! t/\tau$, $f_{\text{sin}}(t)\!=\!g_{\text{sin}}(t)\!=\! \sin \left( \pi t/ 2\tau \right)$, 
and $f^{3}_{1/3}(t)\!=\!g_{1/3}(t)\!=\! t/\tau$. The performance of the functions $f(t)$ and $g(t)$ can now be promptly verified. 
By driving the system by the Hamiltonian $H_{\text{ad}}(t)$ and computing the ergotropy $\Ecal^{\text{ad}}_{\text{{QB}}}(\tau)$ 
of the {QB} at the end of the evolution (at time $t\!=\!\tau$), we can analyze the global stability of the energy transfer process. 
To this end, the Hamiltonian and reduced density matrix used in Eq.~(\ref{Ergotropy}) are given by 
$H_{0}^{\text{{QB}}}\!=\!\hbar \omega \sigma_{z}$ and
$\rho^{\text{ad}}_{\text{{QB}}}(t)\!=\! \text{Tr}_{{\Ccal}}[\rho_{\text{ad}}(t)]$, respectively. The result is shown in Fig.~\ref{Fig-ErgoAdiab-Global}. 
Notice that, for a given total time $\tau$, the {QB} ergotropy is approximately equivalent for any interpolation scheme, with some advantage 
for the linear interpolation in most of regions except the region near $\Omega \tau = 10$, where the interpolation via $f_{1/3}(t)$ and $g_{1/3}(t)$ 
is a better option. 
Concerning stability, all the three interpolations are globally stable at the adiabatic 
time scale. In fact, the power of adiabatic {charger}s is dictated by the energy gap structure of 
$H_{\text{ad}}(t)$, while its stability is achieved whenever the adiabatic dynamics takes place  
and the initial state is encoded in a single eigenstate of $H_{\text{ad}}(t)$~\cite{Santos:19-a,Santos:20c}.

\begin{figure}[t!]
	\centering
	\includegraphics[scale=0.32]{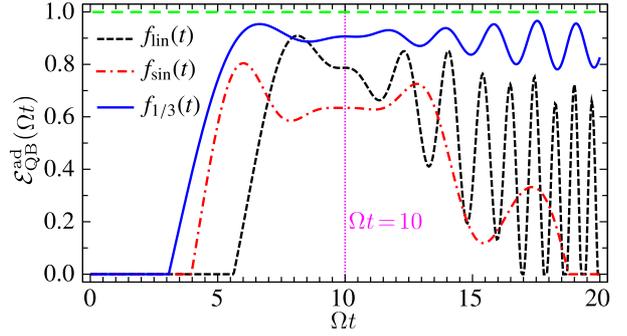}
	\caption{Ergotropy $\Ecal^{\text{ad}}_{\text{{QB}}}(\tau)$ of the {QB} (as a multiple of $\Ecal_{\text{max}}^{\text{qubit}}$) as a function of $t$, for $\Omega\tau\!=\!10$.  
		The horizontal dashed line denotes the maximal 
		charge, which has been normalized to 1.
		Local stability strongly depends on the interpolation function adopted. }
	\label{Fig-ErgoAdiab-Inst}
\end{figure}

We also consider the instantaneous charge as a function of $t$, for a fixed total time $\tau$. This is performed to study the 
local stability for different choices of the interpolating functions. Again, the ergotropy is computed for the {QB} as in 
Eq.~(\ref{Ergotropy}). The results are shown in Fig.~\ref{Fig-ErgoAdiab-Inst}, where we have set $\Omega \tau\!=\!10$. {We highlight the intant in which $\Omega t\!=\!10$ (vertical dotted line), so that we can see the efficiency in the stability of each evolution considered.} 
Notice that the interpolation via $f_{1/3}(t)$ and $g_{1/3}(t)$ is able to asymptotically ensure local stability, since it 
is designed to favor $H_{fin}$ for large $t$. For an always-on connection, 
this interpolation prevents full backflow of energy from the {QB} to the {charger}. 
From the behavior of the ergotropy, it is straightforward to see that the efficiency $\eta_{\text{ls}}(\Delta t)$ can be drastically 
affected by the choice of the interpolating function for $H_{\text{ad}}(t)$.

%%%%%%%%%%%%%%%%%%%%%%%%%%%%%%%%%%%%%%%%%%%%%%%%%
\section{Stability of TQD ergotropy transfer}
%%%%%%%%%%%%%%%%%%%%%%%%%%%%%%%%%%%%%%%%%%%%%%%%%

As previously shown, adiabaticity provides an optimal strategy to achieve global stability, even though local stability will 
depend on the interpolation funtion adopted to drive the system. On the other hand, due to limitations imposed by the 
validity conditions of adiabatic dynamics, we inevitably have a loss of power in the charging/discharging stage. 
In order to provide an alternative approach to recover the high-power performance, keeping stability, we will use 
\textit{transitionless quantum driving} (TQD) for speeding up the energy transfer process.

As originally proposed, the TQD approach for speeding up the adiabatic evolution requires the inclusion of an additional term 
$H_{\text{cd}}(t)$, called \textit{counter-diabatic} Hamiltonian, to the original adiabatic Hamiltonian $H_{\text{ad}}(t)$ so that 
the dynamics is governed by the TQD Hamiltonian $H_{\text{tqd}}(t) = H_{\text{ad}}(t)+ H_{\text{cd}}(t)$, where
\begin{align}
H_{\text{cd}}(t) = i\hbar\sum_{n}\ket{\dot{n}(t)}\bra{n(t)} + \interpro{\dot{n}(t)}{n(t)} \ket{n(t)}\bra{n(t)} , \label{Hcd}
\end{align}
with $\ket{n(t)}$ denoting the set of eigenstates of $H_{\text{ad}}(t)$. Under this consideration, one can speed up the 
adiabatic dynamics by the adiabatic path of $H_{\text{ad}}(t)$, since diabatic transitions are forbidden due to presence 
of the term $H_{\text{cd}}(t)$. It is known that we can arbitrarily speed up the dynamics, given that the $H_{\text{cd}}(t)$ 
can be efficiently implemented (or simulated)~\cite{Santos:15}.

In general, the stability of the battery is not {\it a priori} guaranteed for TQD even in case where we have perfect control 
of the time evolution of $H_{\text{tqd}}(t)$. In fact, the stability of the adiabatic approach is obtained because the system 
state at instant $t\!=\!\tau$ is an eigenstate of $H_{\text{ad}}(t)$, so that the system continuously to evolve to the corresponding 
instantaneous eigenstate at latter times~\cite{Santos:20c}. 
Therefore, we do not need to turn off the {charger}-{QB} coupling. In the TQD case, due to term $H_{\text{cd}}(t)$, 
the state of the system at instant $t\!=\!\tau$ is not eigenstate of $H_{\text{tqd}}(t)$, then some dynamics is expected 
if we do not turn off the {charger}-{QB} interaction. However, in cases where we use the boundary conditions to get 
$H_{\text{cd}}(t\!=\!0)\!=\!H_{\text{cd}}(t\!=\!\tau)\!=\!0$, the stability is recovered. It is known that such boundary condition 
can be obtained for a number of evolutions~\cite{Ibanez:12,Torrontegui:13,Odelin:19}, but here we will adopt a 
milder assumption, with no boundary conditions on time taken. As we shall see, even under this assumption, 
we get considerable degree of stability while speeding up the ergotropy transfer process.

\begin{figure}[t!]
	\subfloat[$f_{\text{lin}}(t)$]{\includegraphics[scale=0.39]{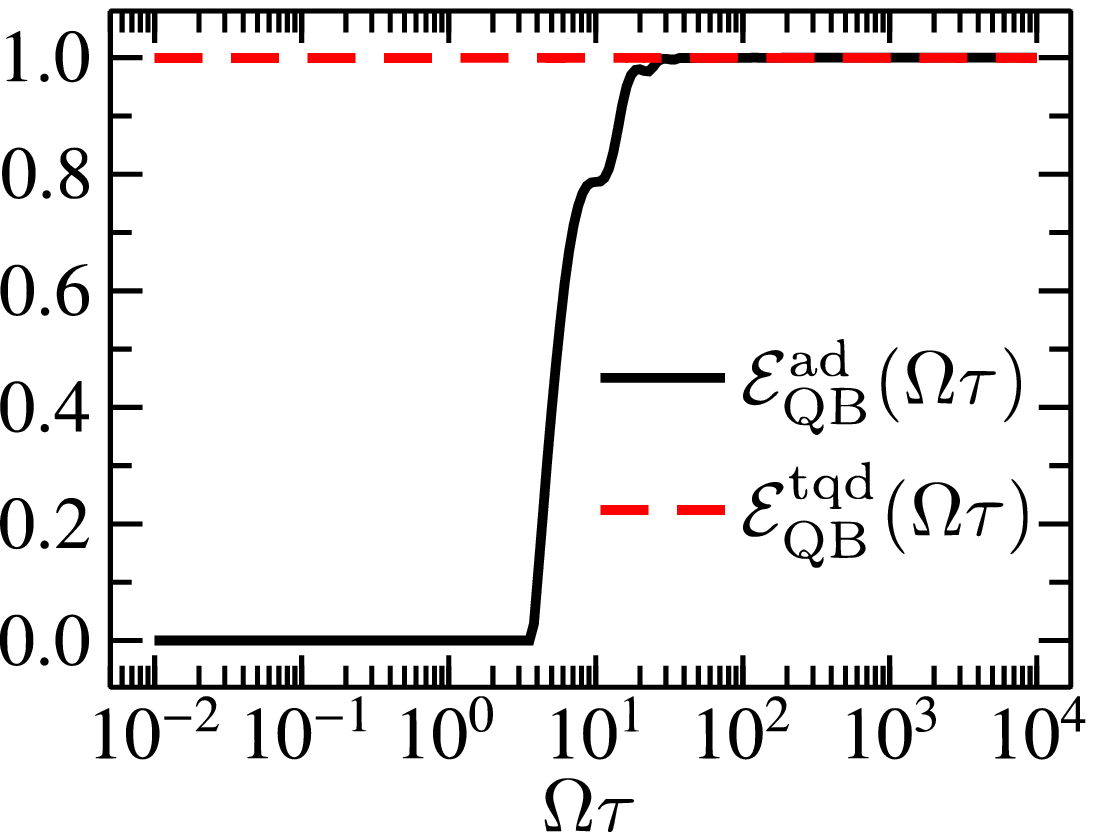}\label{Fig-Super-F1}}~\subfloat[$f_{1/3}(t)$]{\includegraphics[scale=0.39]{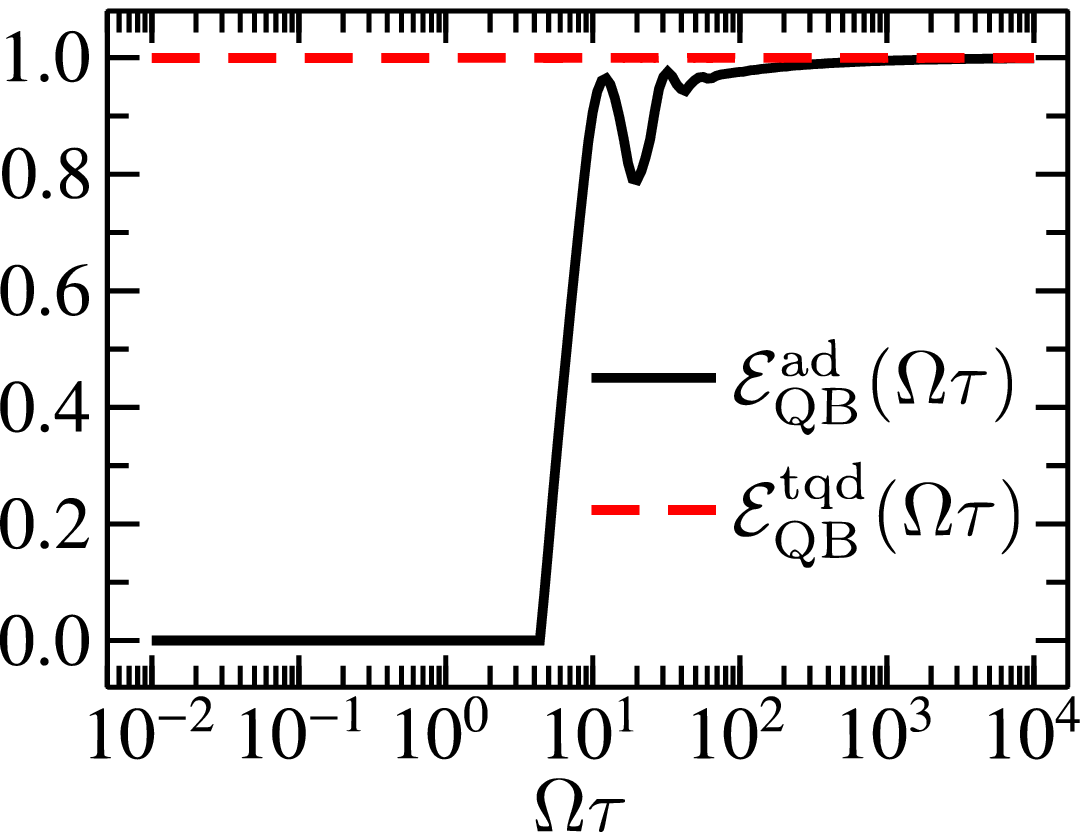}\label{Fig-Super-F3}}
	\caption{Adiabatic and TQD ergotropy transfer from the {charger} to the {QB} (as a multiple of $\Ecal_{\text{max}}^{\text{qubit}}$) at instant $t\!=\!\tau$, as a function of $\Omega\tau$, for 
		distinct interpolation functions, namely, $f_{\text{lin}}(t)$ [Fig. \eqref{Fig-Super-F1}] and $f_{1/3}(t)$ [Fig. \eqref{Fig-Super-F3}]. }\label{Fig1-TQDCharge}
\end{figure}

To study the performance of TQD charging, we consider the shortcut to the adiabatic evolution driven by the Hamiltonian in 
Eq.~\eqref{AdHamil}. First, let us consider global stability. To this end, 
given the counter-diabatic term $H_{\text{cd}}(t)$, we let the {charger}-{QB} evolve under $H_{\text{tqd}}(t)$ 
and then we compute the {QB} ergotropy at $t\!=\!\tau$. In Fig.~\ref{Fig1-TQDCharge} we show the ergotropy 
$\Ecal^{\text{tqd}}_{\text{{QB}}}(\tau)$ of the {QB} after the TQD evolution, where we choose two different 
interpolations, namely, $f_{\text{lin}}(t)$ and $f_{1/3}(t)$ to get a comparison between the adiabatic 
and TQD charging performance. While adiabatic charging requires a specific minimum time $\tau$ to achieve maximum charge 
(or to get some non-zero amount of ergotropy), the counter-diabatic energy transfer leads to a 
maximum charge state at an arbitrary time interval. This is a consequence of the inverse engineering 
approach, where a suitable counter-diabatic contribution allows us to mimic the adiabatic dynamics at 
arbitrary $\tau$, as dictated by the quantum speed limit for counter-diabatic dynamics~\cite{Santos:15}. 
On the other hand, it is worth highlighting that such additional advantage comes at the well-known cost of 
some intensity-field demand in the term $H_{\text{cd}}(t)$ in $H_{\text{tqd}}(t)$~\cite{Santos:15}. 

\begin{figure}[t!]
	\centering
	{\includegraphics[scale=0.30]{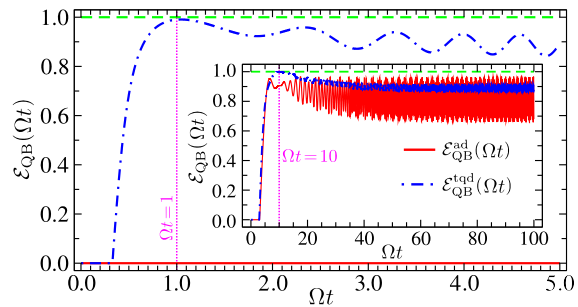}}
	\caption{Adiabatic and TQD ergotropy transfer from the {charger} to the {QB} (as a multiple of $\Ecal_{\text{max}}^{\text{qubit}}$) as a function of $\Omega t$, for the interpolation function 
		given by $f_{1/3}(t)$ e $g_{1/3}(t)$.
		The ergotropy is shown for the case of $\Omega \tau = 1$, where the TQD approach is able to provide a stable charge, while the adiabatic method fails. The horizontal dashed line denotes the maximal 
		charge, which has been normalized to 1. In the inset, a slower evolution is 
		adopted by setting $\Omega \tau\!=\!10$, so that both TQD and adiabatic dynamics achieve local stability.}\label{Fstab}
\end{figure}

Concerning local stability, we can also show that the conter-diabatic approach provides a robust charge 
method even if fluctuations of the total time $\tau$ are taken into account. This is illustrated in Fig.~\ref{Fstab}, 
where both adiabatic and TQD ergotropy transfer from the {charger} to the {QB} (as a multiple of $\Ecal_{\text{max}}$) 
are plotted as a function of $\Omega t$ for different total times $\tau$ and for the interpolation function given by $f_{1/3}(t)$ e $g_{1/3}(t)$. 
Notice that, by adopting $\Omega \tau = 1$, the adiabatic condition is violated. {Again, we highlight the { instant} $\Omega t\!=\!1$ as a vertical dotted line}. In that case, the TQD approach is 
still able to mimic adiabaticity and to provide a locally stable ergotropy transfer. Even though we have the local stability coefficient $\eta_{\textrm{ls}} > 0$, 
the {QB} ergotropy is kept in a large value in the asymptotic regime. In the inset of Fig.~\ref{Fstab}, 
we also consider a case of validity of adiabaticity by setting $\Omega \tau = 10$ {(vertical dotted line denotes the instant $\Omega t\!=\!10$)}. In this situation, both the adiabatic method 
and the TQD approach provide a locally stable charging process from the {charger} to the {QB}.

{
%%%%%%%%%%%%%%%%%%%%%%%%%%%%%%%%%%%%%%%%%%%%%%%%%
\section{Energy cost of the charging process}
%%%%%%%%%%%%%%%%%%%%%%%%%%%%%%%%%%%%%%%%%%%%%%%%%

In this section we briefly discuss the external energy cost of implementing the TQD charging process in { comparison} with the adiabatic { evolution}. 
It is known that the standard approach of TQD is more energetically costly than its adiabatic counterpart~\cite{Santos:15}. However, as we shall see, it is 
possible to find a range of values of the total evolution time { for which} the TQD charging { provides a beneficial setup}. To this end, let us consider 
the measure { $\Sigma(\tau)$} of energy cost { for} implementing a time-dependent Hamiltonian { through}  the interval $t\in[0,\tau]$, { which is provided by}
\begin{align}
\Sigma(\tau) = \frac{1}{\tau} \int_{0}^{\tau} ||H(t)||_{\text{HS}} dt,
\end{align}
where $||A||_{\text{HS}}\!=\!(\tr{A A^{\dagger}})^{1/2}$ is the Hilbert-Schmidt norm of the operator $A$.  { We observe that $\Sigma(\tau)$} has been used in experimental implementations 
of TQD in order to quantify the energy cost of simulating adiabatic and TQD evolutions~\cite{Santos:20b,Hu:18}. { Moreover, it has recently been shown in Ref.~\cite{Deffner:21}} that this 
quantity allows for quantifying the thermodynamic cost of implementing the dynamics driven by an arbitrary Hamiltonian $H(t)$. { By explicitly evaluating the adiabatic and 
TQD energy costs, we then obtain} 
\begin{align}
\Sigma_{\text{ad}} &= \frac{1}{\tau}\int_{0}^{\tau} ||H_{\text{ad}}(t)||_{\text{HS}} dt = \frac{1}{\tau}\int_{0}^{\tau}\sqrt{\sum_{n} E_{n}^{2}(t)} dt , \\
\Sigma_{\text{tqd}} &= \frac{1}{\tau}\int_{0}^{\tau} ||H_{\text{tqd}}(t)||_{\text{HS}} dt \nonumber \\ &= \frac{1}{\tau}\int_{0}^{\tau}\sqrt{\sum_{n} E_{n}^{2}(t) + \hbar^2 \mu_n(t)} dt , \nonumber
\end{align}
where $\mu_n(t)=\interpro{\dot{n}(t)}{\dot{n}(t)} - |\interpro{n(t)}{\dot{n}(t)}|^2$~\cite{Santos:16}. Therefore, given the parameters $\Omega$ and $\tau$, and the functions $f(t)$ and $g(t)$, the quantities above can be 
{ directly} computed. { In our case, we consider the linear function $f_{\text{lin}}(t)$ and perform the integration by numerical methods, with results} shown in Fig.~\ref{EnergyCost}. It is also 
convenient to define the ratio $\Sigma_{\text{rel}} = \Sigma_{\text{tqd}}/\Sigma_{\text{ad}}$, which quantifies how { larger the energy cost is in the TQD dynamics in comparison with} its adiabatic counterpart. 
The result for $\Sigma_{\text{rel}}$ is shown in { the inset of } Fig.~\ref{EnergyCost}.

\begin{figure}[t!]
	\centering
	{\includegraphics[scale=0.34]{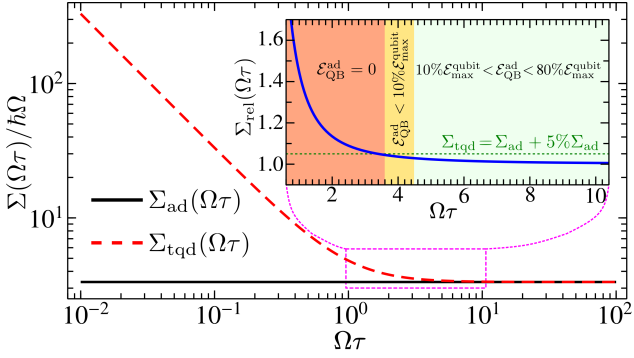}}
	\caption{Energy cost { (as a multiple of $\hbar\Omega$) as a function of  $\Omega\tau$ for the implementation of the adiabatic and TQD charging of the QB. 
	Inset: The relative cost of the TQD process in comparison with the adiabatic charging}. Here we consider the interpolation function $f_{\text{lin}}(t)$.}\label{EnergyCost}
\end{figure}

As expected, the faster the charging process is, the { larger } the energy cost to implement it in a TQD process, as we can see in Fig.~\ref{EnergyCost}. 
However, it is possible to find a range of values for the charging time $\tau$ { for which} the TQD protocol presents { better performance with respect to the adiabatic case}. 
In fact, in the { inset of} Fig.~\ref{EnergyCost}, we consider the time interval between $\Omega\tau$ and $10\Omega\tau$. As previously shown in Fig.~\ref{Fig-Super-F1}, for this interval, the TQD process 
allows us for reaching the full charged state of the battery, while the adiabatic evolution cannot charge the battery efficiently ({ with the maximum being} around $80\%$ of $\Ecal_{\text{max}}^{\text{qubit}}$). 
Then, by taking into account the cost of implementing the transitionless evolution, relatively to the adiabatic { dynamics}, it is worth supporting the charging process via TQD. 
For example, consider the range of values where $\Ecal^{\text{ad}}_{\text{QB}}<10\%\Ecal_{\text{max}}^{\text{qubit}}$, as highlighted in the yellow box in Fig.~\ref{EnergyCost} (inset). For this interval of values for $\Omega\tau$, 
the maximum value for $\Sigma_{\text{tqd}}(\Omega\tau)$ corresponds to { a mild TQD} increasing cost, with $\Sigma_{\text{tqd}}\!=\Sigma_{\text{ad}}+\!5\%\Sigma_{\text{ad}}$.}

%%%%%%%%%%%%%%%%%%%%%%%%%%%%%%%%%%%%%%%%%%%%%%%%%
\section{Conclusion}
%%%%%%%%%%%%%%%%%%%%%%%%%%%%%%%%%%%%%%%%%%%%%%%%%
We have proposed a solution for the local stability of { QBs} as well as for the inherent 
control of their charging power. This allows for the implementations of always-on {charger}s that are able 
to charge a {QB} as fast as the energy available for their internal dynamics. The {charger}-{QB} model proposed here is 
composed by a set of independent three-qubit cells, which ensures the locality of the interaction Hamiltonian. 
The energy resources can be further optimized by implementing a more general approach of TQD technique~\cite{Santos:18-b}. 
In this approach, the phase accompanying the state vector that mimics the adiabatic evolution is set as a free parameter that can be used to adjust the 
interaction Hamiltonian for a smoother TQD energy requirement. This is a certainly interesting point for further analysis. 
Moreover, it is also potentially fruitful the investigation of the relationship between the correlations present in the {charger} and 
its performance with respect to the energy storage and transfer. 
This may bring insights about possible generalizations of the {charger} model to favor the relevant 
correlations. These topics are left for future research. 

{It is worth mentioning that we recently became aware of a work in which the authors also consider TQD for speeding up the adiabatic charging process of three-level quantum batteries~\cite{Hanyuan:21}.}

\acknowledgments

L. F. C. M. acknowledges financial support from the Coordena\c{c}\~ao de Aperfei\c{c}oamento de Pessoal de N\'{\i}vel Superior (CAPES).
A.C.S. acknowledges financial support through the research grant from the S\~ao Paulo Research Foundation (FAPESP) (Grant No 2019/22685-1).
M.S.S. acknowledges financial support from the Conselho Nacional de Desenvolvimento  Científico e Tecnológico  (CNPq) (No. 307854/2020-5).  
This research is also supported in part by CAPES (Finance Code 001) and by the Brazilian National Institute for Science and Technology of Quantum Information [CNPq INCT-IQ (465469/2014-0)].

%\bibliography{mybib-URL}
%\bibliographystyle{eplbib.bst}

\end{document}